\documentclass[prl,twocolumn,showpacs,amsmath,amssymb,superscriptaddress]{revtex4}
\usepackage[dvips]{graphicx}

\usepackage{graphics}

\usepackage{dcolumn}

\usepackage{bm}

\usepackage[usenames]{color}

\begin{document}

\title{Curie point singularity in the temperature
derivative of resistivity in (Ga,Mn)As}

\author{V.~Nov{\'ak}}
\affiliation{Institute of Physics ASCR, v.v.i., Cukrovarnick\'a 10, 162 53 Praha 6, Czech Republic}

\author{K.~Olejn{\'{i}}k}
\affiliation{Institute of Physics ASCR, v.v.i., Cukrovarnick\'a 10, 162 53 Praha 6, Czech Republic}

\author{J.~Wunderlich}
\affiliation{Hitachi Cambridge Laboratory, Cambridge CB3 0HE, United Kingdom}
\affiliation{Institute of Physics ASCR, v.v.i., Cukrovarnick\'a 10, 162 53 Praha 6, Czech Republic}

\author{M.~Cukr}
\affiliation{Institute of Physics ASCR, v.v.i., Cukrovarnick\'a 10, 162 53 Praha 6, Czech Republic}

\author{K.~V\'yborn\'y}
\affiliation{Institute of Physics ASCR, v.v.i., Cukrovarnick\'a 10, 162 53 Praha 6, Czech Republic}

\author{A.~W.~Rushforth}
\affiliation{School of Physics and Astronomy, University of Nottingham, Nottingham NG7 2RD, United Kingdom}

\author{K.~W.~Edmonds}
\affiliation{School of Physics and Astronomy, University of Nottingham, Nottingham NG7 2RD, United Kingdom}

\author{R.~P.~Campion}
\affiliation{School of Physics and Astronomy, University of Nottingham, Nottingham NG7 2RD, United Kingdom}

\author{B.~L.~Gallagher}
\affiliation{School of Physics and Astronomy, University of Nottingham, Nottingham NG7 2RD, United Kingdom}

\author{Jairo~Sinova}
\affiliation{Department of Physics, Texas A\&M University, College
Station, TX 77843-4242, USA}
\affiliation{Institute of Physics ASCR, v.v.i., Cukrovarnick\'a 10, 162 53 Praha 6, Czech Republic}

\author{T.~Jungwirth}
\affiliation{Institute of Physics ASCR, v.v.i., Cukrovarnick\'a 10, 162 53 Praha 6, Czech Republic}
\affiliation{School of Physics and Astronomy, University of Nottingham, Nottingham NG7 2RD, United Kingdom}

\date{\today}

\begin{abstract}
We observe a singularity in the temperature derivative $d\rho/dT$ of resistivity at the Curie point of high-quality (Ga,Mn)As ferromagnetic semiconductors with $T_c$'s ranging from approximately 80 to 185~K. The character of the anomaly is sharply distinct from the critical contribution to transport in conventional dense-moment magnetic semiconductors and is reminiscent of the $d\rho/dT$ singularity in transition metal ferromagnets.
Within the critical region accessible in our experiments, the temperature dependence on the ferromagnetic side can be explained by dominant scattering from uncorrelated spin fluctuations. The singular behavior of  $d\rho/dT$ on the paramagnetic side points to the important role of short-range correlated spin fluctuations.
\end{abstract}

\pacs{75.40.-s,75.50.Pp,5.47.-m}

\maketitle

Since the seminal works of de~Gennes and Friedel \cite{DeGennes:1958_a} and Fisher and Langer \cite{Fisher:1968_a}, critical behavior of resistivity has been one of the central problems in the physics of itinerant ferromagnets. Theories of coherent scattering from long wavelength spin fluctuations, based on the original paper by de~Gennes and Friedel, have been used to explain the large peak in the resistivity $\rho(T)$ at the Curie temperature $T_c$ observed in Eu-chalcogenide magnetic semiconductors \cite{Haas:1970_a}. The emphasis on the long wavelength limit of the spin-spin correlation function, reflecting critical behavior of the magnetic susceptibility,  is justified in these systems by the small carrier density and corresponding small Fermi wavevector.

As pointed out by Fisher and Langer  \cite{Fisher:1968_a}, the resistivity anomaly in high carrier density transition metal ferromagnets is qualitatively different and associated with the critical behavior of correlations between nearby moments. When approaching $T_c$ from above, thermal fluctuations between nearby moments are partially suppressed by short-range magnetic order.  Their singular behavior is like that of the internal energy and unlike that of the magnetic susceptibility. The singularity at $T_c$ occurs in $d\rho/dT$ and is closely related to the critical behavior of the specific heat $c_v$. While Fisher and Langer expected this behavior for $T\rightarrow T_c^+$ and a dominant role of uncorrelated spin fluctuations at $T\rightarrow T_c^-$, later studies of elemental transition metals found a proportionality between $d\rho/dT$ and $c_v$ on both sides of the Curie point \cite{Joynt:1984_a,Shacklette:1974_a}.

In this paper we explore critical contribution to the resistivity of the ferromagnetic semiconductor (Ga,Mn)As. After a decade of intense research, (Ga,Mn)As has become an archetypical carrier-mediated dilute moment ferromagnet. The dilute magnetic semiconductors  have unique micromagnetic and magnetotransport properties \cite{Matsukura:2002_a,Jungwirth:2006_a} and have been utilized in a number of prototype semiconductor spintronic devices \cite{Chiba:2003_a,Yamanouchi:2004_a,Wunderlich:2006_a,Pappert:2007_a,Stolichnov:2008_a}. Nevertheless, the search and proper description of the nature of the critical contribution to resistivity has remained one of the outstanding open problems in the field \cite{VanEsch:1997_a,Matsukura:1998_a,Omiya:2000_a,Kuivalainen:2001_a,Yuldashev:2003_a,Lopez-Sancho:2003_b,Timm:2004_a,Moca:2007_b,Dietl:2007_d}. Recalling the behavior of dense-moment, low carrier density magnetic semiconductors \cite{Haas:1970_a}, long wavelength spin fluctuation effects have been considered to contribute to a broad peak in $\rho(T)$ near $T_c$ observed in higher resistive materials \cite{Matsukura:1998_a,Omiya:2000_a,Kuivalainen:2001_a,Dietl:2007_d}. In annealed, more conductive films the peak further broadens into a shoulder whose dominant temperature dependence on the ferromagnetic side of the transition has been ascribed to scattering from uncorrelated Mn-impurities \cite{VanEsch:1997_a,Lopez-Sancho:2003_b,Moca:2007_b}. An implicit assumption has been commonly adopted that critical scattering is either unimportant or that any sharp singularity at $T_c$ is blurred due to the inherently disordered nature of these heavily doped semiconductors with local magnetic moments randomly distributed in the lattice.

Here we report a strong singularity at $T_c$ in $d\rho/dT$ in thin (Ga,Mn)As epilayers prepared under optimized growth and postgrowth annealing conditions, with nominal Mn-dopings ranging  from 4.5 to 12.5\% and corresponding  Curie temperatures from 81 to 185~K. The nature of the singularity is discussed qualitatively by analyzing  thermodynamic and transport properties  at zero and finite magnetic fields, by monitoring magnetization and resistivity during several successive annealing steps, and  by comparing temperature dependent magnetization and resistivity of optimized materials near $T_c$. We relate our observations to studies of critical behavior in other itinerant ferromagnets and  remark on the relevance of our work for the ongoing  discussion on the nature of ferromagnetism and Curie temperature limit in (Ga,Mn)As materials with increasing Mn content.

The key observation is illustrated in Fig.~1 on measurements in a nominally 12.5\% Mn-doped,  23~nm thick (Ga,Mn)As film with $T_c=185$~K. The material was grown by low-temperature molecular beam epitaxy (MBE) on a rotating semi-insulating GaAs substrate and buffer layer. Growth and post-growth annealing conditions were optimized for the specific Mn-doping and layer thickness to achieve the high Curie temperature and  low concentration of unintentional charge and moment compensating defects. Details of the reproducible preparation techniques for high quality (Ga,Mn)As materials will be described elsewhere. Temperature dependent magnetization and resistivity curves were measured on the same physical specimen of lateral dimensions of $4.5\times 4.5$~mm in one experimental system (Quantum
Design MPMS) equipped with a precise temperature controller and a superconducting quantum interference device (SQUID).
Different sample holders required for the magnetic and transport measurements may have caused a relative shift of the temperature axes in the two experiments which we estimate to be below 1~K. Using the minimum sweep rate of 3~K/min and time averaging we were able to obtain data with the best temperature resolution of 0.5~K, i.e., the accessible range of $t=|T-T_c|/T_c$ is limited in our measurements to values $\gtrsim 3\times 10^{-3}$.

\begin{figure}
\includegraphics[width=1.0\columnwidth,angle=0]{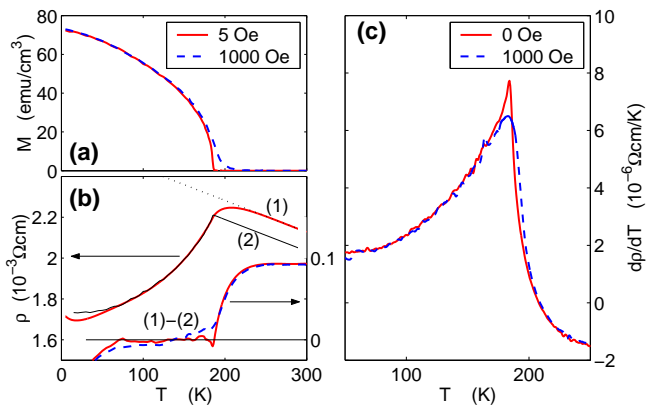}
\caption{
(a) Temperature dependent SQUID magnetization of an optimized  12.5\% Mn-doped (Ga,Mn)As measured at 5
and 1000~Oe.
(b) Measured resistivity $\rho(T)$ at zero-field (1), the fit of $\rho(T)$  by $\rho_{fit}(T)=c_0+c_{nm}T+c_{m2}M^2(T)+c_{m4}M^4(T)$ (2), and the difference $\rho(T)-\rho_{fit}(T)$ for the 0 and 1000~Oe field measurements (right hand scale). For details on the fitting see the main text.
(c) Temperature derivative of the measured resistivity at 0 and 1000~Oe. The peak in zero-field $d\rho/dt$ coincides, within the experimental error, with the SQUID $T_c=185~K$.
}
\label{fig1}
\end{figure}

Fig.~1(a) shows magnetization $M(T)$ measured at 5~Oe and 1000~Oe field along the [1$\bar{1}$0]-direction which is the  easy-axis in this material over the whole studied temperature range. The 5~Oe field was applied to avoid domain reorientations when measuring the temperature dependent remanent magnetization. The remanence vanishes sharply at $T\rightarrow T_c^-$, confirming very good quality of the material. In Fig.~1(b) we show results of corresponding zero-field and 1000~Oe-field measurements of $\rho(T)$. A useful insight into the physics of the shoulder in $\rho(T)$ near $T_c$ is obtained by first removing the non-magnetic part in the temperature dependence, $c_{nm}T$, which we approximate by linearly extrapolating from the high-temperature $\rho(T)$ data (see the straight dotted line in Fig.~1(b)). The measured $M(T)$ then allows us to  subtract the contribution from uncorrelated scattering, assuming a $M^2$ expansion dependence \cite{Fisher:1968_a,VanEsch:1997_a,Lopez-Sancho:2003_b,Moca:2007_b}. We obtained the parameters for the expansion from the best fit of the zero-field $\rho(T)$ to $\rho_{fit}(T)=c_0+c_{nm}T+c_{m2}M^2(T)+c_{m4}M^4(T)$ within the interval from 80~K  to $T_c$, using the remanent $M(T)$. On the ferromagnetic side, a very close fitting can be achieved. On the paramagnetic side, the nose dive of the remaining magnetic contribution to $\rho(T)$ suggests that the shoulder in the measured resistivity originates from a singular behavior  at $T\rightarrow T_c^+$ rather than from disorder broadening effects.

The singularity is revealed by numerically differentiating the experimental zero-field $\rho(T)$ curve, as shown in Fig.~1(c). The position of the sharp peak in $d\rho/dT$ coincides, within experimental error, with the Curie temperature determined from $M(T)$; the peak cut-off correlates with the resolution of our resistivity and temperature measurements. The critical nature of the transport anomaly is confirmed by measurements in the 1000~Oe field: $\rho(T)$ at $T<T_c$ can be closely fitted by $\rho_{fit}(T)$ using the 1000~Oe $M(T)$ and the same values of the fitting constants as in zero field. The field removes the singular behavior at $T_c$ and  makes the onset of the magnetic contribution to $d\rho/dT$ from the paramagnetic side more gradual than in zero field.

We have inspected several tens of materials prepared in two different MBE systems in Prague and Nottingham. We conclude that the $d\rho/dT$ singularity described in Fig.~1 is a generic characteristic of thin (Ga,Mn)As films spanning a wide range of Mn-dopings, prepared by reproducible growth and post-growth annealing procedures which had been optimized separately for each doping level. This is illustrated in  Fig.~2 where we plot $\rho(T)$ and $d\rho/dT$ data measured in optimized 4.5-12.5\% Mn-doped materials. The samples show sharp $d\rho/dT$ singularities at Curie temperatures ranging from  81 to 185~K.

\begin{figure}
\includegraphics[width=1.0\columnwidth,angle=0]{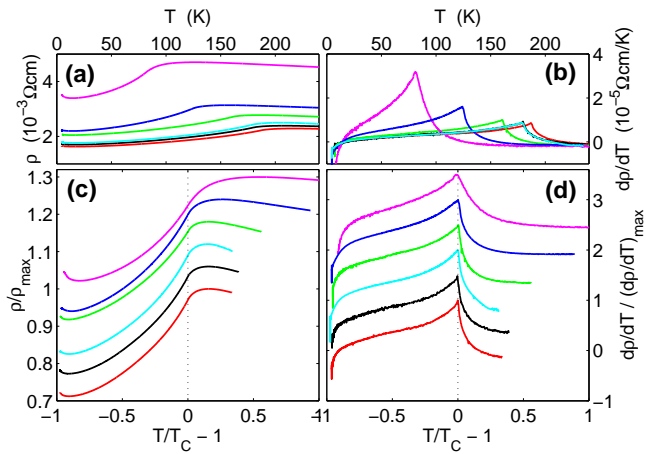}
\caption{(a) Resistivities $\rho(T)$ and (b) temperature derivatives $d\rho/dT$ for optimized (Ga,Mn)As films (of thicknesses between 13 and 33~nm) prepared in Prague and Nottingham MBE-systems. Data normalized to maximum $\rho(T)$  and  $d\rho/dT$ are plotted in (c) and (d), resp. Curves in (a),(c), and (d) are ordered  from top to bottom according to increasing Mn-doping and $T_c$: 4.5\% Mn-doped sample with $T_c=81$~K, 6\% with 124~K, 10\% with 161~K, 12\% with 179~K (Prague and Nottingham samples), and 12.5\% doped sample with 185~K Curie temperature. In (c) and (d) curves are offset for clarity.}
\label{fig2}
\end{figure}

\begin{figure}
\includegraphics[width=0.9\columnwidth,angle=0]{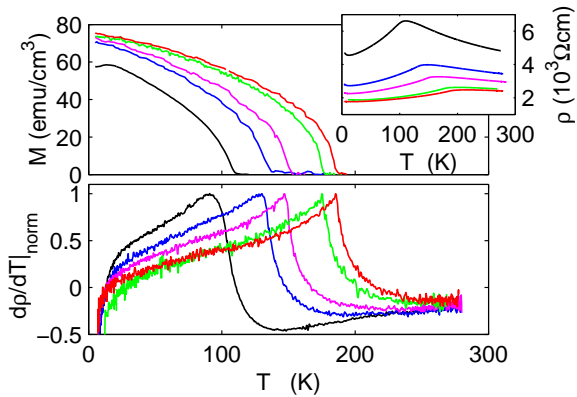}
\caption{
Upper panel: Magnetization curves of one 12.5\% Mn-doped (Ga,Mn)As specimen annealed in
several successive steps. From left to right: as grown, 10, 40, 120 minutes, and 14 hours annealed at 160$^\circ$C in air.
Bottom panel: normalized $d\rho/dT$ curves measured in the same annealing steps.
Resistivity curves are shown in the inset with the highest $\rho(T)$ corresponding to the as-grown state of the sample.
}
\label{fig3}
\end{figure}

The crucial role of material optimization is evidenced in Fig.~3 where we compare magnetization and resistivity curves for annealing sequence measured in one specimen prepared from the 12.5\% Mn-doped 23~nm wafer. We observe a clear correlation between the smeared critical region of $M(T)$ and a slowly increasing $d\rho/dT$ up to the broad peak near $T_c$ in the as-grown film.  The development of sharply vanishing $M(T)$ at $T_c$ and the onset of the singularity in $d\rho/dT$ are also well correlated within the annealing sequence. (The main effect of annealing is the removal of interstitial Mn impurities \cite{Matsukura:2002_a,Jungwirth:2006_a}.) Confirming results were obtained  in a set of samples prepared from a 100~nm thick (Ga,Mn)As wafer and subsequently etched to thicknesses of 50,  and 25~nm. After optimal annealing of each of the three films, highest $T_c$ and sharpest singularity in $d\rho/dT$ of the form shown in Figs.~1 and 2 was achieved in the 25~nm film in which we expect the smallest concentration of residual unintentional impurities. A comprehensive study of deviations from the critical behavior of Figs.~1 and 2, which includes also higher resistive low-doped, or ultra-thin (below 10~nm) (Ga,Mn)As films, will be presented elsewhere. Here we focus on the phenomenology of the singular $d\rho/dT$ curves which we now analyze in more detail.

Ferromagnetism in (Ga,Mn)As originates from 
spin-spin coupling between local Mn-moments and valence band holes, $J\sum_i\delta({\bf r}-{\bf R}_i){\bf s}\cdot{\bf S}_i$ \cite{Matsukura:2002_a,Jungwirth:2006_a}. Here ${\bf S}_i$ represents the local spin  and ${\bf s}$ the hole spin-density operator. This local-itinerant exchange  interaction plays a central role in theories of the critical transport anomaly. When treated in the Born approximation, the interaction yields a carrier scattering rate from magnetic fluctuations which is proportional to the static spin-spin correlation function, $\Gamma({\bf R}_i,T)\sim J^2[\langle{\bf S}_i\cdot{\bf S}_0\rangle - \langle{\bf S}_i\rangle\cdot \langle{\bf S}_0\rangle]$ \cite{DeGennes:1958_a}. Typical temperature dependences of the uncorrelated part, $\Gamma_{uncor}({\bf R}_i,T)\sim \delta_{i,0}J^2[S(S+1) - \langle{\bf S}_i\rangle^2]$, and of the Fourier components, $\Gamma({\bf k},T)=\sum_{i\neq 0}\Gamma({\bf R}_i,T)\exp({\bf k\cdot R}_i)$, are illustrated in the lower inset of Fig.~4 \cite{Fisher:1968_a}. At small wavevectors, $\Gamma({\bf k},T)$ has a peak near $T_c$; at $k$ similar to the inverse separation of the local moments ($kd_{\uparrow - \uparrow}\sim 1$) the peak broadens into a shoulder while the singular behavior at $T_c$ is in the temperature derivative of the spin-spin correlator.

The $M^2$ expansion, providing a good fit to the magnetic contribution to the resistivity $\rho_m(T)=\rho(T)-c_{nm}T$ at $T<T_c$, corresponds to the dominant contribution from  $\Gamma_{uncor}$ on the ferromagnetic side of the transition within experimentally accessible values of $t$. The shoulder in $\rho_m(T)$ and the presence of the singularity in $d\rho_m/dT$ on the paramagnetic side suggests that  large wavevector components of $\Gamma({\bf k},T)$ dominate the temperature dependence of the scattering in the  $T\rightarrow T_c^+$ critical region. This is consistent with the ratio between hole and Mn local-moment densities approaching unity, i.e., with large carrier Fermi wavevectors in optimized (Ga,Mn)As materials \cite{Jungwirth:2005_b}. The property makes (Ga,Mn)As distinct from the dense-moment magnetic semiconductors \cite{Haas:1970_a} and makes its critical transport anomaly reminiscent of elemental transition metal ferromagnets \cite{Joynt:1984_a,Shacklette:1974_a}.

A more detailed assessment of the temperature dependence of the resistivity is obtained from  $\log(d\rho_m/dt)$~vs.~$\log(t)$ plots shown in the right panel of Fig.~4 for the 6-12.5\% doped materials. (We have not included data for the 4.5\% doped sample which are consistent with the other data but have higher experimental noise.) All data sets collapse into a common temperature dependence for $T<T_c$ and another common dependence for $T>T_c$.

\begin{figure}
\includegraphics[width=.8\columnwidth,angle=90]{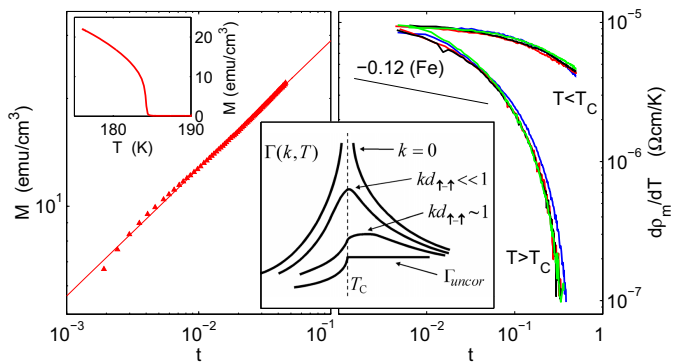}
\caption{
Left panel: $\log(M)-\log(t)$ plot of a $T_c=185$~K sample measured at zero magnetic field. Fitted straight line shows a power-law with an exponent $0.36$. The top inset shows the linear $M(T)$ plot.
Right panel: $\log(d\rho/dT)-\log(t)$ plots  for the 6-12.5\% doped samples of Fig.~2 (horizontally shifted to overlap).
Straight line corresponding to exponent -0.12 ($c_v$ in Fe) is included for comparison.
Lower inset: Schematics of the uncorrelated part and Fourier components  of the spin-spin correlation function for different ratios between wavevector and inverse of the spin separation.}
\label{fig4}
\end{figure}

The initial strength of the temperature dependence of $d\rho_m/dT$ as $T$ approaches $T_c$ from above is large. In iron, e.g.,  $d\rho_m/dT$ follows closely the specific heat, which in the critical region is given by $c_v\sim t^{-\alpha}$ with the exponent $\alpha\approx 0.12$. To qualitatively estimate thermodynamic critical exponents in (Ga,Mn)As we plot in the left panel of Fig.~4 high-accuracy, zero external field $\log(M)-\log(t)$ data for a $T_c=185$~K sample. We can fit the data with a power-law dependence with an approximate exponent $\beta\sim 0.3-0.4$, consistent with either
Heisenberg or Ising behavior for both of which $\alpha\sim 0.1$. (Note that quantitatively accurate determination of critical exponents is a subtle experimental area which is beyond the scope of this paper.) High precision studies of the critical region are necessary to establish to what extent the  stronger initial temperature dependence of $d\rho_m/dT$ we observe reflects a more intriguing relationship between magnetism and transport in (Ga,Mn)As \cite{Timm:2004_a} and whether the proportionality between $d\rho/dT$ and $c_v$ is recovered  in the closer vicinity of $T_c$ \cite{Klein:1996_a,Kim:2003_b}. Within the accessible range of $t$'s we do not observe a clear power-law behavior in $d\rho/dt$ on either side of the transition.

We point out that the common characteristics of transport near $T_c$ in the optimized materials, illustrated in Figs.~2, and 4, bear important implications for studies of the Curie temperature limit in (Ga,Mn)As. The record $T_c=185$~K, 12.5\% Mn-doped material represents our current highest doping for which we achieved reproducible optimization of growth and annealing conditions (which becomes exceedingly tedious for concentrations above 10\%).
Comparing the $d\rho/dT$ singularity in this material with the lower-doped samples provides an experimentally simple yet physically appealing probe which suggests that despite the increase of $T_c$ by more than 100~K the physical nature of the magnetic state has not significantly changed up to this largest Mn-doping studied so far.

To conclude, we have described the phenomenology of the Curie point transport anomaly in (Ga,Mn)As. The shoulder in $\rho(T)$ and the singularity in $d\rho/dT$ at $T_c$, which we observe in optimized materials over a wide range of Mn-dopings and $T_c$'s, can be consistently interpreted in terms of large wavevector scattering of carriers from spin fluctuations. Our results have implications for the basic understanding of magnetism in (Ga,Mn)As and make it possible to study the critical transport anomaly in the  class of ferromagnetic semiconductors which have a much simpler band structure and larger variability of key parameters compared to metal ferromagnets. We have also provided a tool for direct transport measurement of $T_c$ in bulk (Ga,Mn)As and in microdevices in which standard magnetometry is not feasible.


\bigskip

\begin{acknowledgments}
We acknowledge measurements by Vojt\v{e}ch Krej\v{c}i\v{r}\'{\i}k and Jakub Jungwirth, discussions with, Charles Gould and Allan MacDonald, and support from EU IST-015728, Czech Republic AV0Z1-010-914, KAN400100625, LC510,  FON/06/E001, FON/06/E002, and U.S. NRI-SWAN, ONR-N000140610122, and DMR-0547875.
\end{acknowledgments}



\end{document}